\documentclass[%
 reprint,
 amsmath,amssymb,
 aps,
]{revtex4-1}

\usepackage{graphicx}
\usepackage{dcolumn}
\usepackage{bm}

\begin{document}

\preprint{APS/123-QED}

\title{Flowing emulsions through disorder:\\Critical depinning and  smectic rivers}

\author{Marine Le Blay}
 \altaffiliation[Also at ]{Total SA. P\^ole d'Etudes et Recherche de Lacq, BP 47-64170 Lacq, France}
\author{Mokhtar Adda-Bedia}%
\author{Denis Bartolo}%
 \email{denis.bartolo@ens-lyon.fr}
\affiliation{%
 Univ. Lyon, ENS de Lyon, Univ. Claude Bernard,\\ CNRS, Laboratoire de Physique, F-69342, Lyon.}%




\date{\today}

\begin{abstract}
During the past sixty minutes only, oil companies have extracted six trillions liters of oil from the ground, i.e. the volume of about two hundreds Olympic swimming pools. This phenomenal number gives a striking illustration of the impact of multiphase flows on the world economy and environment. 
From a fundamental perspective, we now clearly understand the large-scale patterns formed when liquid interfaces are driven through heterogeneous environments. In stark contrast, the displacement of fragmented fluids through disordered media remains limited to isolated droplets and bubbles. 
Here, we elucidate the collective dynamics of emulsions hydrodynamically driven through disordered environments. Advecting hundreds of thousands of microfluidic droplets through random lattices of pinning sites, we establish that the mobilization of confined emulsions is a critical dynamical transition. Unlike contact-line depinning, emulsion mobilization is not triggered by large-scale avalanches but merely requires the coordinated motion of small groups of particles. Criticality arises from the correlations of  seemingly erratic depinning events over system-spanning scales along smectic river networks. We elucidate the microscopic origin of these self-organized flow patterns: contact interactions and hydrodynamic focusing conspire to mobilize emulsion out of disorder. We close our article commenting on the similarities (and profound differences) with the plastic depinning transitions of driven flux lines in high-$T_{\rm c}$ superconductors, and grain transport in eroded sand beds.
\end{abstract}

\maketitle


\section{\label{sec:level1}Introduction}

From enhanced oil recovery to water remediation and $CO_2$ sequestration, a number of industrial processes having a prominent impact on the world economy and environment rely on multiphase flows in heterogeneous media. 
Since the early days of oil production, at the beginning of the twentieth century, engineers noticed that the displacements of confined immiscible fluids result in inhomogeneous spatial patterns limiting oil extraction from rocks and soils~\cite{Buckley1942}.
Since then, these patterns have been extensively studied by fluid mechanicians, statistical physicists and engineers. However, from a fundamental perspective, our understanding has remained strongly unbalanced.
The imbibition and drainage patterns formed by moving interfaces separating two immiscible fluids have been extensively investigated~\cite{Homsy87,Lenormand1990,Sahimi1993,Sandnes2011,Levache2014,Juanes2016,Odier2017}. Conversely, the dynamics of fragmented interfaces in foams and emulsions remains virtually uncharted, although relevant to a number of practical situations~\cite{Payatakes_Review,Diphare2014,Perazzo2018}.
Understanding the displacement of say oil in water emulsions in a natural porous medium is a formidable challenge that remains out of reach of current physical investigations. 
The problem requires addressing at once the transport, the coalescence, the fragmentation, and the coarsening of the droplets, modelling multiscale flows and their interplay with the droplets'shape and speed~\cite{Datta2013,Alim2017}. Until now, aside from rare exceptions, most physics studies have therefore focused on the microscopic mechanisms occuring at the pore scale, or on flows past isolated obstacles~\cite{Bremond2008,Protiere2010,Datta2014,Geraud2016,Dollet2017,Yeates2019}.

Here, in the spirit of the seminal work by Saffman and Taylor, we introduce a minimal microfluidic experiment to address the collective transport properties of particles hydrodynamically driven through heterogeneous environments. 
Considering the advection of monodisperse emulsions past random lattices of pinning sites, we demonstrate a sharp dynamical transition between a creeping regime where droplets undergo finite displacements, and a flowing regime where a finite fraction of the emulsion is effectively mobilized through static and sparse river patterns. 
We establish the critical nature of the mobilization transition and elucidate the smectic symmetry of the resulting flow patterns. 
Here criticality originates from the self-organized geometry of river networks algebraically correlated over system spanning scales. Unlike the elastic depinning of meniscii and contact lines~\cite{Alava2004,Le2009,Snoeijer2013} mobilization does not rely on scale-free avalanches, or large-scale dynamical heterogeneities.  
We conclude discussing the universality of plastic flows in disordered media, and identify the essential similarities (and difference) between hydrodynamic mobilization, plastic depinning in dirty superconductors~\cite{Reichhardt_Review}, granular erosion~\cite{Yan2016,Aussillous2016} and Darcy flows in yield-stress fluids~\cite{Rosso2019,Waisbord2019}.

\section{Experiments in patterned microfluidics}
\begin{figure*}
\includegraphics[width=\textwidth]{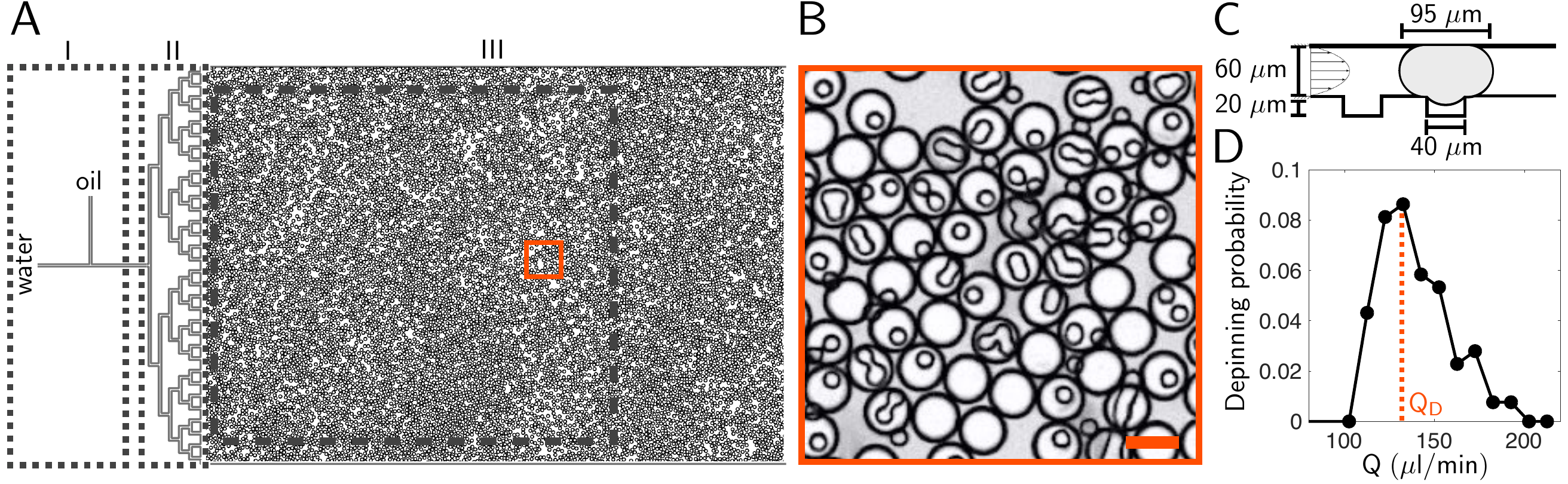}
\caption{Microfluidic experiment. (A) Sketch of the microfluidic device composed of three modules. I: a T-junction is used to produce monodisperse droplets of hexadecane in a water-glycerol solution. II: a homogenizer module splits the droplet stream into 32 parallel channels. III: the main microfluidic channel is fed with the monodisperse emulsion. It includes a random lattice of overlapping circular wells of diameter $40\,\sf \mu m$. We perform our measurement in the window delimited by the dashed line. (B) Close-up on a $800\,{\sf \mu m}\times 800 \, \mu \sf m$ window showing pinned and free droplets. Scale bar: $100\,\sf \mu m$. (C) Sketch of a squeezed droplet pinned by cylindrical well. Side view. (D) PDF of the depinning flow rates measured on a dilute ensemble of monodisperse droplets. The distribution is peaked on the typical value $Q_{\rm D}= 120\,\sf\mu l/min$.
}
\label{Fig1}
\end{figure*}

Our experimental setup is sketched in Fig.~\ref{Fig1}A, see also Methods. We form a monodisperse emulsion of hexadecane at a T-junction in an aqueous solution of Glycerol (40 wt\%), surfactant (Sodium Dodecyl Sulfate (SDS) 0.1 wt\%) and fluorescent dye.
The SDS surfactant prevents coalescence and wetting of the channel walls by the hexadecane droplets. The resulting stable emulsion is injected in a $4.5\,\rm cm$ long and $2\,\rm cm$ wide Hele-Shaw cell through a homogenizer module. In the main channel of height $60\,\rm \mu m$ the drops are squeezed and have a pancake shape of radius $a=47\,\rm \mu m$, see Figs.~\ref{Fig1}B and~\ref{Fig1}C. %
Each experiment is initialized injecting three pore volumes of the emulsion at a solvent flow rate $Q=15\,\rm \mu l/min$, see Supplementary Video 1. When the injection stops, we are left with a uniform distribution of droplets with an area fraction of 0.7 for all experiments reported in the main text.
Disorder is introduced by patterning the bottom wall of the main channel with circular wells of radius $20\,\rm \mu m$ and depth $20\,\rm \mu m$. As extensively studied in~\cite{Dangla2011}, when a droplet contacts a well, it deforms, relaxes its surface energy, and therefore gets pinned by the well, Figs.~\ref{Fig1}B and~\ref{Fig1}C. In our experiments the pinning-well centers are uncorrelated in space, they are distributed according to a Poisson distribution with a mean density $\rho_{\rm p}=160\,\rm mm^{-2}$. In other words, on average, there are $1.01$ available pinning site per droplet. 
As the wells can overlap (see Fig.~\ref{Fig1}B) we quantify the disorder strength by measuring the critical depinning flow speed associated with each pinning site. To do so we inject a dilute fraction of droplets in the disordered channel and continuously increase the flow rate. The distribution plotted in Fig.~\ref{Fig1}D defines a typical depinning flow rate $Q_{\rm D}=120\, \rm\mu l/min$, for isolated droplets. 
$Q_{\rm D}$ corresponds to a capillary number $Ca \sim 2\times 10^{-3}$. 

\section{Droplet mobilization as a dynamical phase transition} 
\begin{figure*}[t!]
\includegraphics[width=1\textwidth]{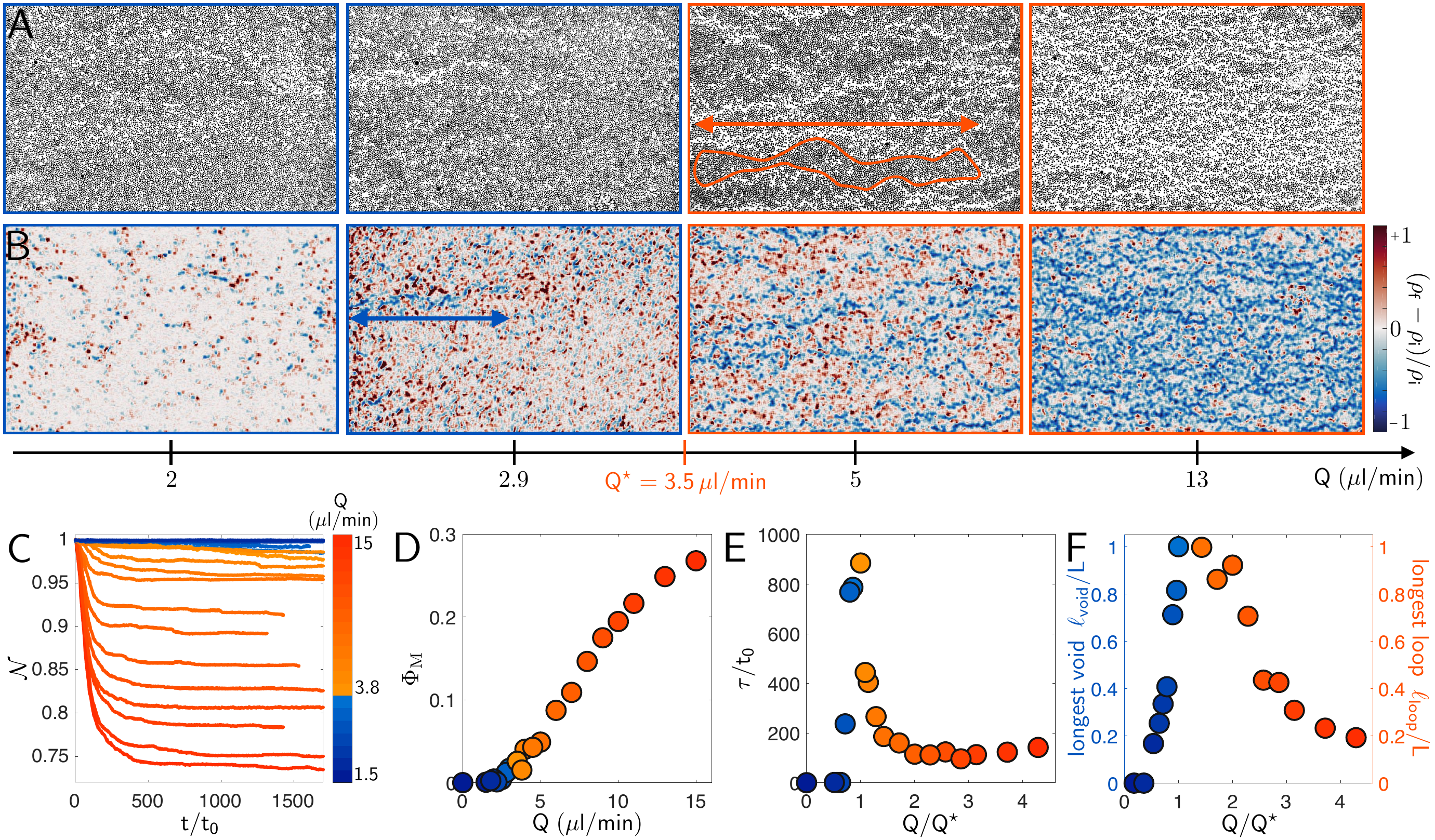}
\caption{The mobilization of emulsion through a disordered microfluidic channel is a critical dynamical transition. (A) Images of the droplets in the disordered channel after the injection of three pore volumes at increasing flow rates $Q$. (B) The color map indicates the difference between the local droplet density at the end and at the beginning of the experiments. Red (resp. blue) indicates a local increase (resp. decrease) of the density. Same experiments as in (A). (C) Variations of the overall droplet fraction $\mathcal N$ as a function of time. Time is normalized by $t_0$, the typical time it takes for a free droplet to move over its own radius. (D) Fraction of droplets mobilized out of the system plotted as a function of the flow rate. $\Phi_{M}$ bifurcates at $Q^\star=3.5\,\mu \sf l/min$. (E) The normalized relaxation time of $\mathcal N(t)$ defined as $\tau/t_0=(1/t_0)\int_0^\infty \! [{\mathcal N}(t)-{\mathcal N}(\infty)]{\rm d}t$ peaks at $Q=Q^\star$. (F) Blue dots: length of the longest depleted void measured when $Q<Q^\star$, see (B). Orange dots: length of the longest loop formed in the river Network, see (A) and {\it SI Appendix}. Both length scales grow nonlinearly with $Q$ and peak at the onset of mobilization.}

\label{Fig2}
\end{figure*}
We now describe the mobilization of concentrated emulsions. The experiment consists in driving at constant flow rate an emulsion trapped by the lattice of pinning site, see supplementary Movies 2 and 3. We stress that no additional droplet is produced in the course of the mobilization experiment. Tracking the individual trajectories of hundreds of thousands oil droplets, we distinguish two qualitatively different regimes illustrated in Fig.~\ref{Fig2}A and Supplementary Videos 3 and 4. Creeping regime: at low flow rates, the droplets undergo finite displacements but ultimately remain trapped in the disordered channel. In Fig.~\ref{Fig2}A showing the state of the system after the injection of three pore volumes, we merely observe the formation of locally depleted regions. Comparing the final and initial density fields, we find that only a small fraction of droplets are displaced, and only over a finite distance, thereby forming droplet depleted regions Fig.~\ref{Fig2}B. Increasing the flow rate, increases the area of the depleted regions and the number of finite-displacements events. Mobilization regime: Above $Q^\star=3.5\,\rm \mu l/min$, macroscopic patterns emerge, Fig.~\ref{Fig2}A. Virtually all droplets contribute to the formation of a sparse network of branched and reconnected rivers percolating throughout the entire system, Fig.~\ref{Fig2}A and ~\ref{Fig2}B. Remarkably, $Q^\star$ is 30 times smaller than the individual depinning threshold $Q_{\rm D}$ which reveals the cooperative nature of the mobilization transition.

The mobilization patterns reflect the net transport of the particles. In Figs.~\ref{Fig2}C we plot the variations of the number of droplets ${\mathcal N}(t)$ as a function of time. In the creeping regime, almost no droplet is extracted. Conversely, in the mobilization regime a finite fraction of the droplets escapes the channel. 
The transition between the two dynamics is sharp. Inspecting the variation of the fraction of mobilized droplets, $\Phi_{\rm M}$, as a function of the applied flow (Figs.~\ref{Fig2}D), we find that $\Phi_{\rm M}$ bifurcates from zero to a finite value when $Q$ reaches $Q^\star$. In order to distinguish between a sharp crossover and a genuine dynamical transition, we first plot the time ($\tau$) needed to reach a stationary state in Fig.~\ref{Fig2}E. Importantly, we compare the different time scales normalizing $\tau$ by the time ($t_0(Q)$) it takes for a free droplet to be advected over a distance $a$ at the applied flow rate. In the creeping regime $\tau/t_0$ is vanishingly small as droplets hardly move over a diameter. Conversely, deep in the mobilization regime $\tau$ converges towards the free advection time over the entire observation window. However, the variations of $\tau$ are non-monotonic and sharply increase in the vicinity of $Q^\star$ where $\tau/t_0$ is three orders of magnitude larger than in the creeping regime. This behavior is suggestive of critical slowing down. To further confirm the hypothesis of a critical scenario, we measure the variations of two typical length scales associated with the mobilization patterns: the length $\ell_{\rm void}$ of the largest depleted void in the creeping regime, and the size of the largest loop $\ell_{\rm loop}$, viz the maximal reconnection length, of the river patterns in the mobilization regime, see Fig.~\ref{Fig2}A and~\ref{Fig2}B. We show in Fig.~\ref{Fig2}F that both length scales diverge at the onset of mobilization. Altogether, the sharp bifurcation of $\Phi_{\rm M}$, the divergence of the relaxation time and the divergence of the length scales characteristic of the rivers' geometry establish the critical nature of the mobilization transition. \\

\section{Critical flow rate}
\begin{figure*}[t!]
\centerline{
\includegraphics[width=\textwidth]{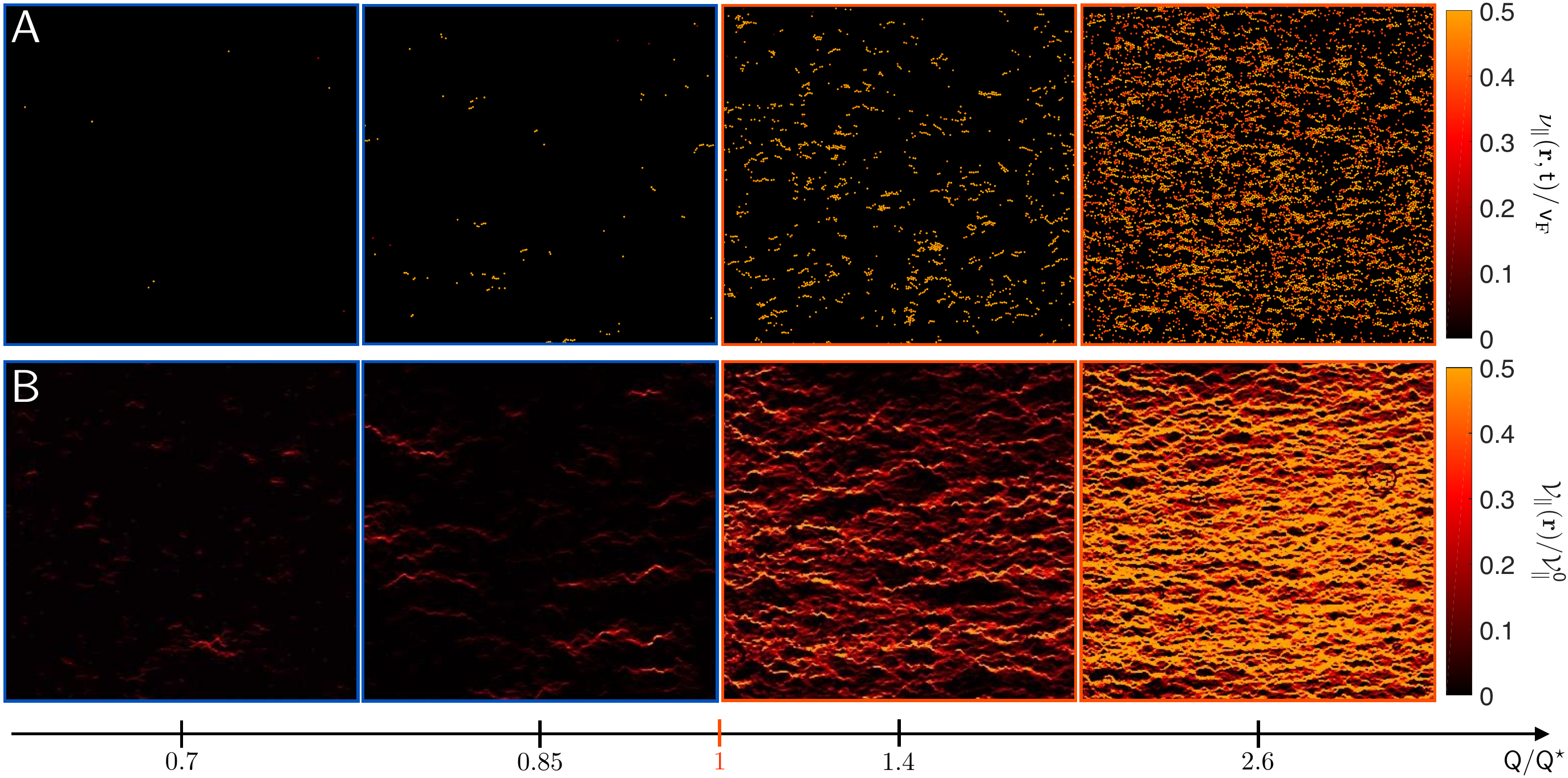}
}
\caption{The droplets are mobilized through a smectic network of channels. (A) Instantaneous droplet-velocity field. The color indicates the magnitude of the longitudinal component $\nu_\parallel(\mathbf r,t)$ evaluated at the time where the slope of the $\mathcal N(t)$ curve is minimum (Fig.~\ref{Fig2}C). (B) Map of the maximal droplet flux $\mathcal V_\parallel(\mathbf r)$: a filamentous river network forms above $Q^\star$.}

\label{Fig3}
\end{figure*}
To gain more insight we now investigate the migration dynamics of the droplets. Fig.~\ref{Fig3}A shows an instantaneous map of the longitudinal component of the droplet-velocity field $\nu_\parallel(\mathbf r,t)$, when the droplet extraction rate is maximal. In both regimes, droplet motion takes the form of localized bursts of activity involving clusters of droplets at contact, see also Supplementary Movies 2, 3, 4 and 5. These localized events seem homogeneously distributed in space, reflecting the nucleation and intermittent growth of depleted regions across the whole sample. Unlike elastic depinning of liquid interfaces, or of elastic lattices, the plastic flows at the onset of mobilization are not supported by large scale avalanches. As a matter of fact, the typical number of droplets in the moving clusters is comprised between two and eight and does not diverge at $Q^\star$, see {\it SI Appendix}. Criticality does not stem from scale-free dynamical heterogeneities.  

The finite size of the local flow bursts, however, reveals that the value of the critical flow rate $Q^\star$ is determined both by lubricated contact interactions and by the geometrical focusing of the solvent flow. The depinning of an isolated droplet is driven by the pressure drop over its own diameter~\cite{Dangla2011}. At the transition, when a cluster of a typical size of five droplet diameters forms along the longitudinal direction, the pressure drop across whole cluster is $\sim5$ times larger than the pressure drop across an isolated particle. One would therefore expect $Q^{\star}$ to be five times smaller that $Q_{\rm D}$ due to the (lubricated) contact interactions which is about an order of magnitude larger than the measured value. This discrepancy is solved noting that droplets at rest form a porous structure that locally focuses the flow of the incompressible continuous phase. The finite porosity of the static emulsion amplifies the local flow rate by a factor of the order of $\sim1/0.3$, where 0.3 is the area fraction of the region accessible to the driving fluid at $t=0$. Altogether hydrodynamic focusing and contact interactions give an estimate of $Q^\star$ of the order of $\sim 7\,\rm \mu l/min$ in qualitative agreement with our measurements ($Q^\star=3.5\,\rm \mu l/min$). We expect this overestimate to reflect the exponentially distributed flow heterogeneities reported in similar porous structures~\cite{Alim2017}.\\

\section{A scale-free smectic network}
We now characterize and elucidate the geometry of the river pattern. To do so, we first define the maximal droplet flux ${\mathcal V_\parallel}(\mathbf r)=\max\left[\nu_\parallel (\mathbf r,t)\right]$ which continuously bifurcates from zero to a finite value at $Q^\star$, when averaged over space, see Fig.~\ref{Fig4}A. Remarkably, the spatial fluctuations of ${\mathcal V_\parallel}(\mathbf r)$ reveal that the seemingly erratic activity bursts seen in Fig.~\ref{Fig3}A are actually strongly correlated in space. A direct comparison between Figs.~\ref{Fig2}A and~\ref{Fig3}B indicates that all depinning events occur along the sparse river networks seen in the final images of our experiments. From a Lagrangian perspective, this dynamics translates into a bimodal distribution of the individual droplet speed $\nu_i$ normalized by the average flow speed $v_{\rm F}$,  Fig.~\ref{Fig4}B. 
The localization of the droplet dynamics explains why only thirty percent of them can be effectively extracted out of the device, even at the highest flow rates. 

In order to quantitatively characterize the geometry of the river network, we compute the two point correlations of the ${\mathcal V_\parallel}$ field, $C_{\mathcal V}(x,y)$. The spatial variations of $C_{\mathcal V}(x,y)$ exhibit three essential features:
(i) The velocity correlations decay exponentially with marked oscillations along the transverse direction, Fig.~\ref{Fig4}C. 
The oscillations reflect the periodicity of the river pattern along the $y$ axis with a period $\lambda=2a$ irrespective of the magnitude of the flow rate. Remarkably, the emergence of this translational order coincides with the onset of mobilization. We demonstrate this structural transition by plotting the magnitude of the oscillations, $S=C_{\mathcal V}(0,2a)-C_{\mathcal V}(0,a)$, as a function of the flow rate in Fig.~\ref{Fig4}D, and show that $S$ continuously bifurcates from zero to a finite value at $Q^\star$. 
(ii) The correlations of $\mathcal V_\parallel$ decay algebraically along the flow direction: $C_{\mathcal V}(x,0)\sim x^{-\alpha}$ with $0.5<\alpha<1$, Fig.~\ref{Fig4}E. The smectic river patterns are therefore critical over the entire mobilization regime.  
(iii) The 2D variations of $C_{\mathcal V}(x,y)$ hint towards a microscopic explanation for the rivers' geometry. The oscillatory decay of $C_{\mathcal V}(x,y)$ indeed persists only when the distance vector $\mathbf r$ makes an angle larger than $\pi/4$ with respect to the longitudinal axis, see~Fig.\ref{Fig4}F. 
This angle corresponds to the direction where a dipolar flow perturbation changes its sign along the longitudinal axis thereby suggesting the following picture. 

\section{Dipolar interactions}
\begin{figure*}[t!]
\centerline{
\includegraphics[width=\textwidth]{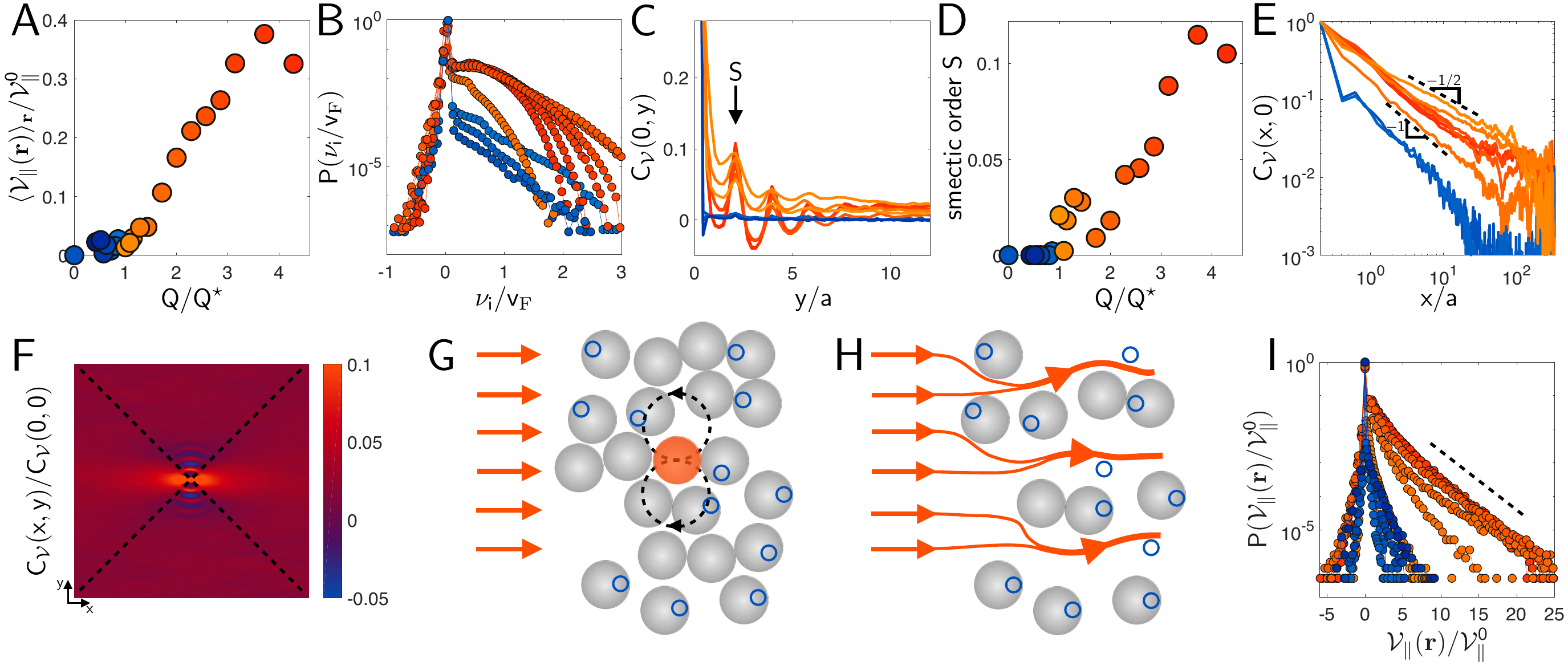}
}
\caption{
(A) Spatial average of the maximal droplet flux $\langle \mathcal V_\parallel(\mathbf r)\rangle_{\mathbf r}$ plotted versus $Q$. The droplet flux $\langle \mathcal V_\parallel(\mathbf r)\rangle_{\mathbf r}$ bifurcates at $Q^\star$. (B) Histogram of the individual droplet speed (along the longitudinal direction). The bimodal distribution reveals the formation of sparse flowing rivers as in~\cite{Faleski96}.
(C) Decorrelations of the longitudinal flux $\langle \mathcal V_\parallel(\mathbf r)\rangle_{\mathbf r}$ along the direction $y$ transverse to the flow. (D) The smectic order parameter $S=C_{\mathcal V}(0,2a)-C_{\mathcal V}(0,a)$ bifurcates at $Q=Q^\star$. Nematic rivers of width $2a$ emerge at the onset of mobilization. (E) Along the longitudinal $x$-direction the correlations of the droplet flux decay algebraically. $C_{\mathcal V}(x,0)\sim x^{-\alpha}$ with $0.5<\alpha<1$. (F) 2D variations of the flux correlations: $C_{\mathcal V}(x,y)$. The dashed lines indicates the directions making an angle of $\pi/4$ and $3\pi/4$ with the $x$-axis. The suppression of the oscillations past the dashed lines hints towards the formation of dipolar force chains. (G) Sketch of the dipolar displacements induced by the depinning of one droplet in a contact cluster. (H) The focusing of the driving flow along the low density regions promotes and stabilizes mobilization along filamentous rivers. (I) Distribution of the instantaneous and droplet flux $\nu_\parallel(\mathbf r,t)$ measured at the time where the decay rate of $\mathcal N(t)$ is maximal. The local droplet current is distributed according to an exponential distribution. This measurement contrasts with the power law reported in erosion experiments and simulations~\cite{Yan2016,Aussillous2016}.  
}
\label{Fig4}
\end{figure*}
Consider the initial configuration sketched in Fig.~\ref{Fig4}G. The depinning of the orange droplet would result in a dipolar displacements field for the surrounding droplets, if they were all free to move. This generic behavior~\cite{Sadhu2011} follows from mass conservation and therefore applies to the quasistatic deformations, or force distribution in any ensemble of nearly incompressible particles, see e.g.~\cite{Candelier2009} for a neat example in a granular medium. Mass conservation implies that the divergence of the particle current vanishes and, at lowest order in gradients, a local particle displacement along the direction $\hat{\mathbf p}$ at the origin corresponds to a localized dipolar perturbation: $\nabla\cdot\mathbf j(\mathbf r)\sim\hat{\mathbf p}\cdot\bm \nabla\delta(\mathbf r)$. The resulting displacements share the same symmetry and are screened over a distance comparable to a particle size (algebraically, or exponentially)~\cite{Demery2014,Morris2016,Poncet2017}. The consequences of this dipolar perturbation are illustrated in Fig.~\ref{Fig4}H. Particles in the downstream region are further pulled thereby promoting the formation of a directed river. Conversely the depinning of the particles located above and below the moving one is hindered by the dipolar contact-force network pushing in the direction opposite to the driving flow. Further away, in the transverse direction, the dipolar perturbation vanishes and the driving fluid can freely contribute to particle depinning. All together these contact mechanisms promote, on average, the formation of river lanes along the flow direction. The formation of a river locally increases the permeability of the medium and therefore facilitates the fluid flow along this anisotropic region of space. This hydrodynamic focusing mechanism stabilizes the rivers and hence prevents them from shrinking or widening. 
In conclusion, the coordinated action of lubricated contact interactions and hydrodynamic focusing explains the self-organization of the droplet flow into a directed river patterns\footnote{It is also worth noting that the spatial correlations of the longitudinal current are reminiscent of the algebraic correlations found in ensembles of interacting particles driven out of equilibrium by constant forces applied to a finite fraction of the particles~\cite{Poncet2017,Lowen2012}}.

\section{Discussion}
The mobilization of emulsions in our disordered microfludic channels realizes a prototypical example of a plastic depinning transition. Plastic depinning was introduced first in the context of flux-line transport in high-$T_{\rm c}$ superconductors~\cite{Watson1996,Troyanovski99,Pardo98,Faleski96,Reichhardt_Review} and further extended to driven soft matter~\cite{Yan2016,Aussillous2016,Pertsinidis2008,Tierno2018,Juniper2016,Reichhardt_Review}. Simply put, it defines a flow transition in ensembles of interacting units driven by a constant force through a disordered landscape of pinning sites. Regardless of the microscopic nature of the driven units, plastic depinning is a dynamical phase transition separating an absorbing static state, and a flowing state where a finite fraction of the particles are driven through filamentous patterns~\cite{Reichhardt_Review}. Although experiments remain scarce, and no unified theory exists, a number of quantitative simulations provide a consistent phenomenology~\cite{Reichhardt_Review,Fily2010}: (i) in 2D the transition is critical~\cite{Watson1996,Reichhardt_Review,Yan2016}, (ii) plastic flows are associated with a bimodal speed distribution~\cite{Faleski96,Pertsinidis2008}, (iii) the instantaneous flows occur along sparse networks of interconnected rivers having a smectic structure deep in the flowing regime~\cite{Balents98,Fily2010,Reichhardt_Review}. Our experiments enjoy these three distinctive properties. 

However, two noticeable differences exist between conventional plastic depinning and emulsion mobilization. Firstly, numerical simulations and mean-field models predict an algebraic flux distribution of the plastic flows qualitatively confirmed by granular erosion experiments~\cite{Yan2016,Aussillous2016}. In contrast, the distribution of the emulsion flux is clearly exponential in Fig.~\ref{Fig4}I. More importantly, in ''dry'' plastic depinning, the river networks continuously rearrange their structure in time, ultimately involving the motion of all particles~\cite{Faleski96,Aussillous2016}. The patterns shown in Figs.~\ref{Fig2}A,~\ref{Fig3}B and Supplementary Movie 3 undergo a markedly different dynamics. Once formed the river network is frozen in time and merely allows a subset of droplets to explore a finite fraction of space: both the creeping and the mobilization regimes are absorbing states for a macroscopic fraction of the driven particles. 

We argue that these essential differences stem from the interplay between the hydrodynamic drive and the pattern geometry. Unlike conventional plastic flows, emulsion mobilization is not the response to a constant drive. As a void forms, it locally increases the permeability of the medium, focuses the flow of the driving fluid and locally amplifies the local force on the arrested droplets which are more easily mobilized. The formation of the river network is therefore a self-organization process reminiscent to the Laplacian growth of viscous fingers~\cite{Homsy87}. Similarly, the self-focusing of the driving flow results in static viscous fingers with a characteristic dendritic geometry that cannot undergo any structural transformation. This picture is further confirmed by wet granular experiments and simulations that resemble to our setup in the limit of zero pinning strength~\cite{Kudrolli2004,Mahadevan2012,Kudrolli2016}, where the hydrodynamic drive results in channeling geometries akin to Laplacian patterns.

In conclusion, we have established that the critical mobilization of emulsions defines a ''universality class'' that relies on two essential ingredients: (i) The drive and the interactions must conspire to allow particles to overcome a local mobilization threshold, and (ii) the magnitude of the driving force must be amplified by local mobilization events. Beyond the specifics of our experiments, this conclusion is supported by recent numerical simulations and experiments on the displacement of yield stress fluids in a model porous media~\cite{Rosso2019,Waisbord2019}, a system which combines the two necessary ingredients for critical mobilization and which displays a strikingly similar self-organized plastic depinning phenomenology~\cite{Rosso2019}.

\section{Conclusion}

The collective mobilization of droplets trapped by random lattices of pinning sites results in large-scale patterns, chiefly governed by the interplay between hydrodynamic flows, contact interactions and capillary pinning force. We therefore expect these generic mechanisms to shape the emergent flow patterns of a broad class of fragmented liquid interfaces driven through confined heterogeneous environments. Beyond the specifics of emulsions, we have established that the hydrodynamic mobilization of soft particles in disordered geometries realizes a rare experimental demonstration of a critical plastic depinning transition. We now need to address the robustness of this collective dynamics to strongly nonlinear interactions between particles and disorder such as coalescence and fragmentation, a formidable yet necessary challenge to elucidate the general transport rules of soft matter in heterogeneous media.

\begin{acknowledgments}
We thank Alexandre Morin and Celeste Odier for invaluable help with the experiments, and Enric Santanach-Carreras for insightful discussions. We acknowledge discussions with Alberto Rosso, Cristina Marchetti and Matthieu Wyart.
\end{acknowledgments}

\appendix

\section{Microfabrication}
The microfluidic devices are made using the microfludic-sticker method~\cite{Bartolo2008}. In brief, the fabrication of the stickers is based on the soft imprint of a photocurable resin (NOA 81 Norland). We first make a mold made of two layers of SU8 photoresist (Microchem). The first layer includes the replica of the random lattice of circular wells, the second is a replica of the straight channels. The mold is replicated to make
a PDMS stamp. The stamp is used to imprint a thin layer of thiolene-based resin (NOA 81) cast on a glass coverslip. We cure the resin with a 7s UV exposure ($20\,\rm mW/cm^2$, UV lamp Hamamatsu LC8). The microfluidic chip is finally assembled sealing the sticker with a Quartz slide. Adhesion is ensured by an additional UV exposure and thermal curing for 12h at$ 90^\circ$C. In order to prevent the wetting of the channel walls by the emulsion droplets, we treat the device as described in~\cite{Levache2012}. A deep-UV exposure for 30 mins in a Jelight UVO CLeaner 42 makes the NOA 81 surface permanently hydrophilic. The stickers are connected to three injection tubings using home-made connectors.

\section{Microfluidics and imaging}

We use the stickers to make and transport a hexadecane emulsion dispersed in a solution of water, Glycerol ($40\,\rm wt\%$), surfactant (Sodium Dodecyl Sulfate (SDS) $0.1\,\rm wt\%$) and fluorescein ($0.2\,\rm wt\%$).
To ensure reproducible initial conditions, we use a systematic fluid-injection protocol. First, the sticker is filled with $CO_2$ that is more soluble in water than air. We then fill the whole device with water-glycerol solution and dissolve all $CO_2$ bubbles. Finally, using precision syringe pumps (Nemesis, Cetoni), the water-glycerol solution and hexadecane flow rates are respectively fixed to $Q_w=12\,\rm \mu l/min$ and $Q_h=3\,\rm \mu l/min$ at the T-junction. 
In turn, we produce a monodisperse hexadecane in water-glycerol emulsion with a droplet radius of $a=47 \, \mu m$, and a surface fraction in the main channel of 0.7.
Once the main channel is filled with the emulsion, we stop the hexadecane flow and impose the minimum flow rate possible for the water-glycerol mixture and let the system to relax during ten minutes. The experiment and recording start after imposing the desired flow rate to the aqueous phase.  

We record the fluorescence images of the main channel with a Nikon AZ100 using a 6X magnification and a 4 MPix CCD camera (Basler Aviator). In order to correctly track the droplet trajectories, the frame rate is set as follows: 3.3 fps for flow rates lower than $Q=6\,\rm \mu l/min$, and 10 fps for flow rates higher than $Q=6\,\rm \mu l/min$.\\

\section{Detection and data analysis}

We detect the center of mass of all droplets in the field of view using the ImageJ minima intensity detection function on contrast-enhanced and gaussian blurred frames (the standard deviation of the Gaussian kernel is taken equal to $a/2$)~\cite{ImageJ}. The droplets are detected with a pixel accuracy. We reconstruct the drop trajectories and measure their instantaneous velocity using the MATLAB function by Blair and Dufresne based on the Crocker and Grier tracking algorithm~\cite{Crocker1996}. We compute all the Eulerian and Lagrangian quantities defined in the main text (density, velocity, current fields, droplet speed, etc) from the droplet instantaneous positions and velocities as detailed in {\it SI Appendix}.


%

\end{document}